\documentclass[fleqn,usenatbib]{mnras}

\usepackage{newtxtext,newtxmath}

\usepackage[T1]{fontenc}
\usepackage{ae,aecompl}

\usepackage{graphicx}	
\usepackage{amsmath}	
\usepackage{amssymb}	

\usepackage{float}
\usepackage[figure,figure*]{hypcap}
\usepackage{subfigure}
\usepackage{soul}

\title[DeepStreaks: identifying FMOs in ZTF data]{DeepStreaks: identifying fast-moving objects in the Zwicky Transient Facility data with deep learning}

\author[D. A. Duev et al.]{Dmitry A. Duev,$^{1}$\thanks{E-mail: duev@caltech.edu (DAD)}
    Ashish Mahabal,$^{1}$
    Quanzhi Ye,$^{1,2}$
    Kushal Tirumala,$^{3}$
    \newauthor
    Justin Belicki,$^{4}$
    Richard Dekany,$^{4}$
    S.~Frederick,$^{5}$
    Matthew J. Graham,$^{1}$
    \newauthor
    George Helou,$^{2}$
    Russ R. Laher,$^{2}$
    Frank J. Masci,$^{2}$
    Thomas A. Prince,$^{1}$
    \newauthor
    Reed Riddle,$^{4}$
    Philippe Rosnet,$^{6}$
    Maayane T. Soumagnac,$^{7}$
\\
$^{1}$Division of Physics, Mathematics, and Astronomy, California Institute of Technology, Pasadena, CA 91125, USA\\
$^{2}$IPAC, California Institute of Technology, MS 100-22, Pasadena, CA 91125, USA\\
$^{3}$Division of Engineering and Applied Science, California Institute of Technology, Pasadena, CA 91125, USA\\
$^{4}$Caltech Optical Observatories, California Institute of Technology, Pasadena, CA 91125, USA\\
$^{5}$Department of Astronomy, University of Maryland, College Park, MD 20742, USA\\
$^{6}$Universit\'e Clermont Auvergne, CNRS/IN2P3, LPC, Clermont-Ferrand, France\\
$^{7}$Benoziyo Center for Astrophysics, Weizmann Institute of Science, Rehovot, Israel
}

\date{Accepted 2019 April 13. Received 2019  April 11; in original form 2019 February 25}

\pubyear{2019}

\begin{document}
\label{firstpage}
\pagerange{\pageref{firstpage}--\pageref{lastpage}}
\maketitle

\begin{abstract}

We present \texttt{DeepStreaks}, a convolutional-neural-network, deep-learning system designed to efficiently identify streaking fast-moving near-Earth objects that are detected in the data of the Zwicky Transient Facility (ZTF), a wide-field, time-domain survey using a dedicated 47 deg$^2$ camera attached to the Samuel Oschin 48-inch Telescope at the Palomar Observatory in California, United States. The system demonstrates a 96-98\% true positive rate, depending on the night, while keeping the false positive rate below 1\%. The sensitivity of \texttt{DeepStreaks} is quantified by the performance on the test data sets as well as using known near-Earth objects observed by ZTF. The system is deployed and adapted for usage within the ZTF Solar-System framework and has significantly reduced human involvement in the streak identification process, from several hours to typically under 10 minutes per day.

\end{abstract}

\begin{keywords}
methods: data analysis -- asteroids: general -- surveys
\end{keywords}



\section{Introduction and context}

Solar System small bodies (SBs) in the context of orbiting asteroids and comets are believed to be remnants of our Solar System's early days, holding clues about its formation and evolution. A subclass of SBs known as the near-Earth objects (NEOs) is of particular interest especially since some of them pose a hazard due to a non-negligible probability of collision with the Earth \citep{2013A&A...554A..32D}.\footnote{Of about 19,600 NEOs known as of January 2019, roughly 1,900 are classified as potentially hazardous asteroids (PHAs).} Luckily, collisions with kilometer-sized objects that would have devastating effects are rare. However, the impact frequencies for smaller objects that could still cause significant damage are much higher.

Our knowledge of the kilometer-sized NEO population is fairly complete. However, the current NEO completeness for 140 m objects is only about 30\% and drops rapidly with decreasing object size \citep{2017AJ....154...12V}.\footnote{In 2005, the United States Congress directed NASA to find at least 90 percent of potentially hazardous NEOs sized 140 meters or larger by the end of 2020.} To date, only a relatively small number of NEOs with sizes of 50 m have been discovered, but the vast majority, as many as 98\%, of the estimated quarter million 50 m class NEOs, have not been found yet.

Detection of small NEOs poses a significant challenge as they are either very faint while far away from the Earth, or they have high apparent motion when close and bright enough to be detected. Objects that approach the Earth within $15$ lunar distances typically move at a rate of $>10^\circ$ per day \citep{Verevs2012}. These ``Fast-Moving Objects'' (FMOs) would trail on typical survey exposures (usually 20--60 seconds) and present a challenge for traditional NEO detection algorithms that are most efficient in detecting objects moving slower than a couple degrees per day \citep{Jedicke2013}.

Throughout the paper, we are using the term NEO to refer to the objects of natural origin, confirmed in one way or another, whereas the term FMO is used to refer to the objects that move at a rate of faster than $10^\circ$ per day and could be either natural or human-made.

\subsection*{The Zwicky Transient Facility}

\begin{figure*}
  \centering
  \includegraphics[width=0.9\textwidth]{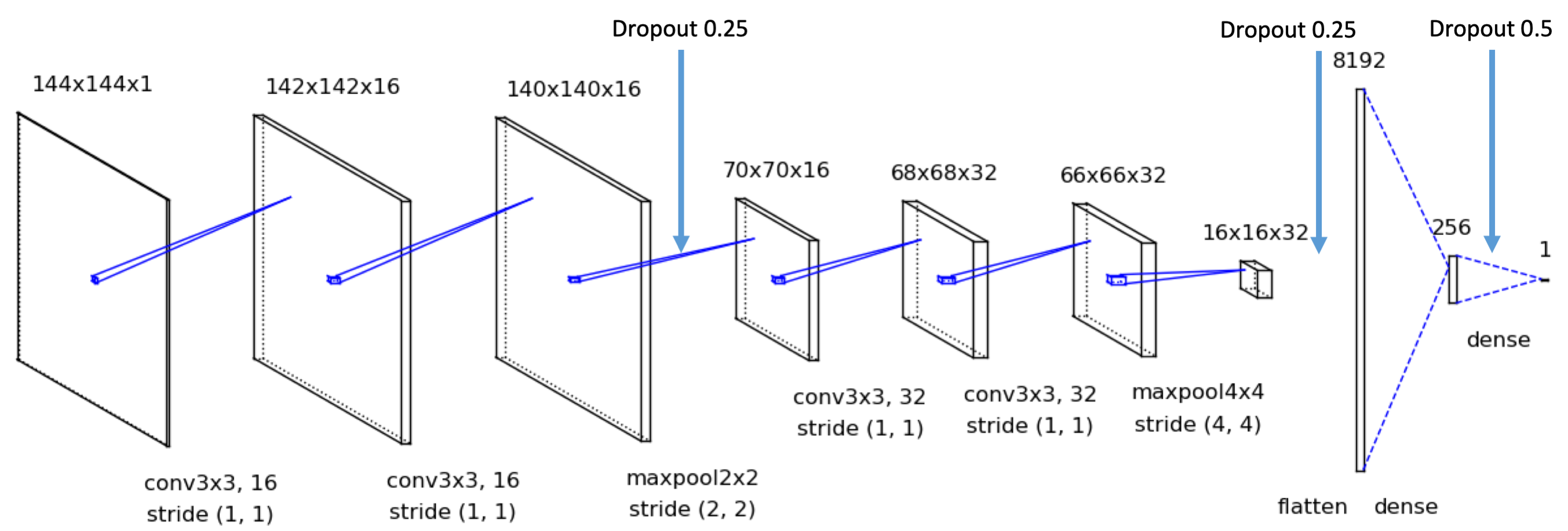}
    \caption{Architecture of the simple custom VGG6 model. ReLU activation functions are used for all five hidden trainable layers; a sigmoid activation function is used for the output layer. Dropout is used for regularization.}
    \label{fig:vgg}
\end{figure*}

The Zwicky Transient Facility (ZTF) is a new robotic time-domain sky survey that visits the entire visible sky north of $-30^\circ$ declination every three nights in the $g$ and $r$ bands, and at higher cadences in selected sky regions including observations with the $i$-band filter \citep{2019PASP..131a8002B, 2019arXiv190201945G}. The new 576 megapixel camera with a 47 deg$^2$ field of view \citep{Dekany2019}, installed on the Samuel Oschin 48-inch (1.2-m) Schmidt Telescope, can scan more than 3750 deg$^2$ per hour, to a $5\sigma$ detection limit of 20.7 mag in the $r$ band with a 30-second exposure during new moon.

The raw data are transferred to the Infrared Processing and Analysis Center (IPAC) at the California Institute of Technology (Caltech) and processed in real time. The ZTF Science Data System (ZSDS) housed at IPAC consists of the data processing pipelines, data archives, infrastructure for long-term curation, and the services for data retrieval and visualization \citep{2019PASP..131a8003M}.

A part of the ZSDS, the ZTF Streak pipeline (ZStreak) focuses on the detection of streaked objects. A detailed description of ZStreak is given in \citet{ye2019zstreak}, and is based, in part, on earlier prototype work by \citet{Waszczak2017}. In essence, the pipeline first detects plausible streak candidates in difference images\footnote{Constructed by ``properly'' subtracting a reference (template) image from a science exposure image according to \citet{2016ApJ...830...27Z}.} by searching for contiguous bright pixels that exceed a signal-to-noise threshold of 1.5 sigma and whose spatial distribution is approximately linear according to an estimate of the Pearson correlation coefficient\footnote{Currently, alternative streak detection approaches are being investigated, including those described in \citet{Nir2018}}. It then tries to fit a streaked point-spread function (PSF). If successful, a manually selected set of features per streak is passed through a Random-Forest (RF) machine-learning (ML) classifier that assigns a score from zero to one representing the likelihood of the streak being real (which corresponds to a score of one). A threshold of 0.05 is adopted, which is about $96-98\%$ complete at this score in terms of detecting real FMOs present in the raw-streak sample. The candidates passing this threshold are vetted for real detections by human scanners on a daily basis. The detected real streaks are linked and if plausible linkages are found, an orbit fit is attempted. Finally, if the orbital solution converges and the corresponding ``track'' is not associated with human-made objects, the observations (of both known and newly discovered objects) are submitted to the International Astronomical Union's Minor Planet Center (MPC)\footnote{\url{https://www.minorplanetcenter.net/}}, the standard clearinghouse for identification, designation and orbit computation for small bodies.

On a typical night, the number of detected ``raw'' streaks (prior to ML classification) reaches $10^5$ -- $10^6$. The RF classifier initially used in production only reduces this to $10^4$ -- $10^5$ still resulting in several human-hours spent on candidate scanning each day. Furthermore, given the number of streaks that need to be looked at, it is not uncommon for the human scanners to miss several streaks from real FMOs. Typically, only several streaks to a few dozen are marked nightly as plausible real candidates.

The aim of this work is to build an ML system that has a sensitivity similar to that of the RF classifier, but significantly reduces the number of false positives.

\section{DeepStreaks: a Deep Learning framework for streak identification}

The term ``artificial intelligence'' (AI) usually refers to situations where machines solve problems commonly associated with human intelligence, such as image recognition and classification. Machine learning, often recognized as a subset of AI, refers to the cases where machines learn from the data rather than being explicitly programmed. Finally, Deep Learning (DL) is a subset of ML that employs many-layer artificial neural networks.

DL has gained popularity in recent years thanks to the advances in both related hardware (graphical and tensor processing units -- GPUs and TPUs) and software (programming frameworks such as \texttt{TensorFlow}, \texttt{PyTorch} and others) coupled with the advent of Big Data. As a result, DL-based systems are starting to outperform humans in a number of areas. In particular, a subclass of DL systems, the convolutional neural networks (CNNs), have been demonstrated to yield outstanding performance in image recognition and classification tasks. For a thorough introduction of CNNs refer to, for example, \citet{2018arXiv180710912L} and references therein.

In this paper, we present \texttt{DeepStreaks}, a CNN-based deep learning framework developed to efficiently identify streaking FMOs in the ZTF data.

\subsection{Architecture}
\label{sec:architecture}

\begin{figure}
  \centering
  \includegraphics[width=0.45\textwidth]{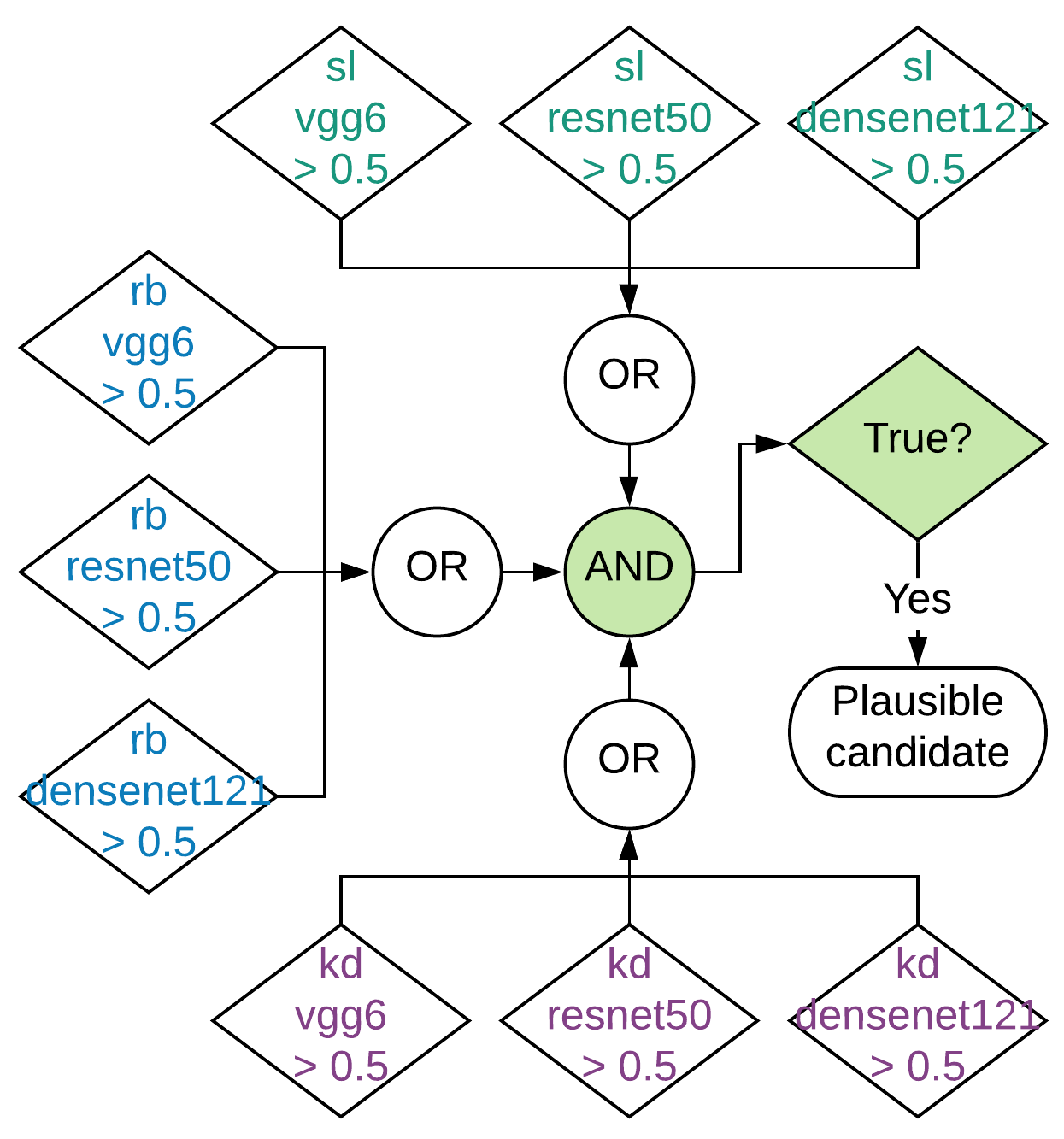}
    \caption{Decision logic used by \texttt{DeepStreaks} to identify plausible streaks. The problem is split into three simpler sub-problems, each solved by a dedicated group of classifiers assigning real vs. bogus (``rb''), short vs. long (``sl''), and keep vs. ditch (``kd'') scores. At least one member of each group must output a score that passes a pre-defined threshold. See Section \ref{sec:architecture} for details.}
    \label{fig:ds.architecture}
\end{figure}

Given the very practical goal of this work, we have chosen to explore two possible DL-system architectures. In the first approach, the problem of identifying a plausible streak is divided into three simpler sub-problems that are each solved by a dedicated group of classifiers:

\begin{enumerate}
    \item \textbf{``rb''}: identify all streak-like objects, including the actual streaks from fast-moving objects, long(er) streaks from satellites, and cosmic rays. Assign a real($rb=1$)/bogus($rb=0$) score.

    \item \textbf{``sl''}: identify short streak-like objects, including the actual streaks from fast moving objects and artifacts such as cosmic rays. Assign a short($sl=1$)/long($sl=0$) score.

    \item \textbf{``kd''}: identify real streaks produced by FMOs. Assign a keep($kd=1$)/ditch($kd=0$) score.

\end{enumerate}

We note that the overwhelming majority of ``long'' streaks are produced by human-made objects. A streak is considered long if it extends outside of the cutout image\footnote{By design, at least some part of the streak is near the center of the cutout image.}. This comprises objects that move faster than  125-175 degrees per day, depending on the streak positional angle.

\begin{figure}
  \centering
  \includegraphics[width=0.35\textwidth]{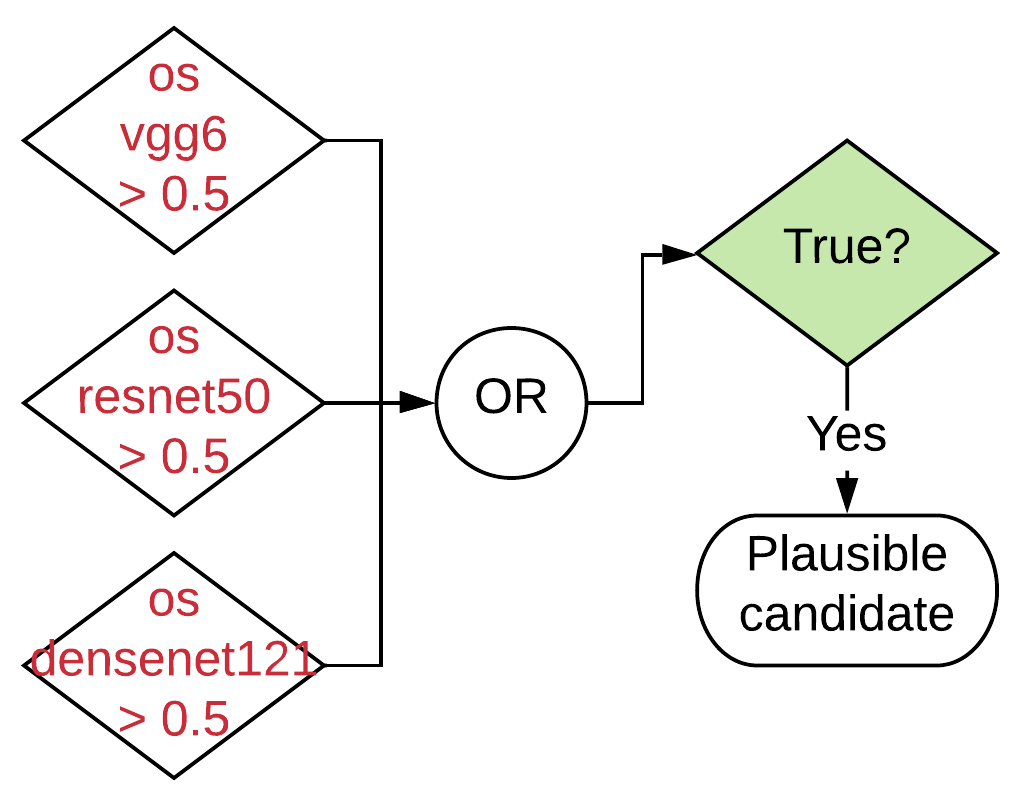}
    \caption{Decision logic to identify plausible streaks used in the one-shot (``os'') classification approach. See Section \ref{sec:architecture} for details.}
    \label{fig:os.architecture}
\end{figure}

The classifiers used in the second, one-shot (\textbf{``os''}) approach that we explored, solve the classification problem directly.

Within each group of classifiers we have chosen to use three different CNN models:

\begin{enumerate}
    \item A simple custom VGG\footnote{This architecture was first proposed by the Visual Geometry Group of the Department of Engineering Science, University of Oxford, UK}-like sequential model (``VGG6'') \citep{2014arXiv1409.1556S} (see Fig. \ref{fig:vgg} for details). The model has six layers with trainable parameters: four convolutional and two fully-connected. The first two convolutional layers use 16 filters each while in the second pair, 32 filters are used. To prevent over-fitting, a dropout rate of 0.25 is applied after each max-pooling layer and a dropout rate of 0.5 is applied after the second fully connected layer. ReLU activation functions\footnote{Rectified Linear Unit --  a function defined as the positive part of its argument} are used for all five hidden trainable layers; a sigmoid activation function is used for the output layer.

    \item A custom 50-layer deep model based on residual connections (``ResNet50''), which are connections that add modifications with each layer, rather than completely changing the signal. The implementation details are given in \citet{2015arXiv151203385H}. As a regularization technique to prevent over-fitting, batch normalization is used.

    \item A custom 121-layer deep model based on dense connections (``DenseNet121''), one of the state-of-the-art models in the field. The implementation details are given in \citet{2016arXiv160806993H}. Similar to ResNet50, batch normalization is used as a regularization technique.
\end{enumerate}

We theorized that an ensemble of relatively shallow and deep CNNs will yield a better classification performance, provided that all models demonstrate high accuracy, since the sets of features extracted and learned by these models will differ dramatically. The performance of individual classifiers will be described in Section \ref{sec:training_performance}.

Architectures that are more complex than ResNet50 and DenseNet121 have been demonstrated to yield better performance on large public image data sets such as ImageNet \citep{imagenet_cvpr09}, however, these models are not necessarily better at generalizing to other data sets \citep{2018arXiv180508974K}.

The differenced cutout images with raw streaks produced by the ZStreak pipeline are gray-scale and of size 157 by 157 pixels (or smaller if a raw streak is detected close to the field edge) at a plate scale of $1\arcsec$ per pixel. The input image size of all our CNN models is 144x144x1, so the cutouts are down-sampled accordingly. All individual models are evaluated on all raw streaks. In the first approach (``rb''+``sl''+``kd''), for a streak to be marked as a plausible real candidate, for each classifier group, at least one group member must output a score greater than a pre-set threshold (see Figure \ref{fig:ds.architecture} and Section \ref{sec:training_performance}). Similarly, in the ``os'' approach, at least one classifier must report a score that passes a threshold (see Figure \ref{fig:os.architecture} and Section \ref{sec:training_performance}).

\subsection{Data sets, training, and performance}\label{sec:training_performance}

To accelerate data labelling, we developed a simple web-based tool we called \texttt{Zwickyverse}\footnote{\url{https://github.com/dmitryduev/zwickyverse}} that provides both efficient API and GUI. The tool is easy to deploy thanks to containerization using \texttt{Docker} software\footnote{\url{https://www.docker.com/}} and it allows quick integration of newly-labelled data sets into the model training workflow. All data labelling for this work was done using \texttt{Zwickyverse}.

\begin{figure*}
  \centering
  \subfigure[Bad subtraction]{\includegraphics[width=0.15\textwidth]{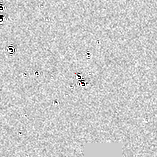}}\quad
  \subfigure[Cosmic ray]{\includegraphics[width=0.15\textwidth]{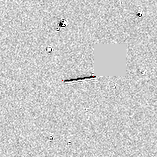}}\quad
  \subfigure[``Dementor'']{\includegraphics[width=0.15\textwidth]{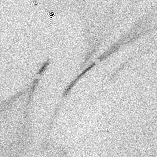}}\quad
  \subfigure[``Ghost'']{\includegraphics[width=0.15\textwidth]{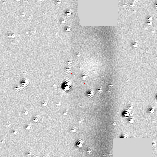}}\quad
  \subfigure[Masked star]{\includegraphics[width=0.15\textwidth]{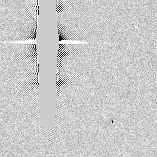}}\quad
  \subfigure[Satellite trail]{\includegraphics[width=0.15\textwidth]{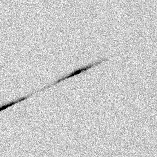}}
\caption{Examples of different classes of bogus streak detections.}
\label{fig:bogus}
\end{figure*}

We started with a training set that consisted of  1,000 differenced images with streaks that span a period of time from the start of the survey in March 2018 until the end of 2018, identified as real by human scanners; 8,270 synthetic streak images generated by implanting a streaked PSF into a bogus image following \citet{ye2019zstreak}; and 6,000 ``bogus'' images of different kinds: streaks from satellites and airplanes (which are typically long and, frequently, of varying brightness) and false streak detections caused by, for example, masked bright stars, bad subtractions, cosmic rays, ``dementors'', and ``ghosts'' (see Figure \ref{fig:bogus}). This data set was used to train an initial set of ``rb'' classifiers that separate all sorts of streak-like objects from false detections. Next, we evaluated the resulting classifiers on a month of ZTF data, sampled images that received low, intermediate, and high scores, labelled the data, and re-trained the classifiers making use of these data as well. This process was repeated several times; the same approach was later applied to all other classifier families.

Next, we trained the ``sl'' classifiers intended to filter out streaks caused by satellites and airplanes, which make up the majority of all streak-like objects in the ZTF data. For that purpose, we used the streaks from the initial data set together with a set of images that received high ``rb'' scores. Finally, ``kd'' classifiers were trained using real and synthetic streaks and a set of cutout images with both high ``rb'' and ``sl'' scores, which is dominated by cosmic rays.

The resulting set of classifiers was deployed in test mode. We then carried out a number of training set ``enrichment'' and classifier retraining campaigns aimed at covering a wider range of weather conditions and tuning the classifier performance. We plan to continue conducting such campaigns in the future.

Separately, the training data set for the ``os'' (one-shot) classifiers contains all true short streaks detected by ZTF from the start of the survey until the end of 2018, the synthetic streaks from the initial data set, and a set of images covering the whole range of possible false streak and bogus images.

As of February 2019, the training set for the ``rb'' classifiers contains 11,857 images of streak-like objects (including the actual streaks from FMOs, long(er) streaks from satellites, and cosmic rays) and 13,449 non-streak images; for the ``sl'' classifiers -- 5,168 long and 11,246  short streak images; for the ``kd'' classifiers -- 14,154 ``false'' and 10,621 ``true'' images; and finally for the ``os'' classifiers -- 16,808 ``false'' and 10,621 ``true'' images.

\begin{figure*}
  \centering
  \includegraphics[width=0.999\textwidth]{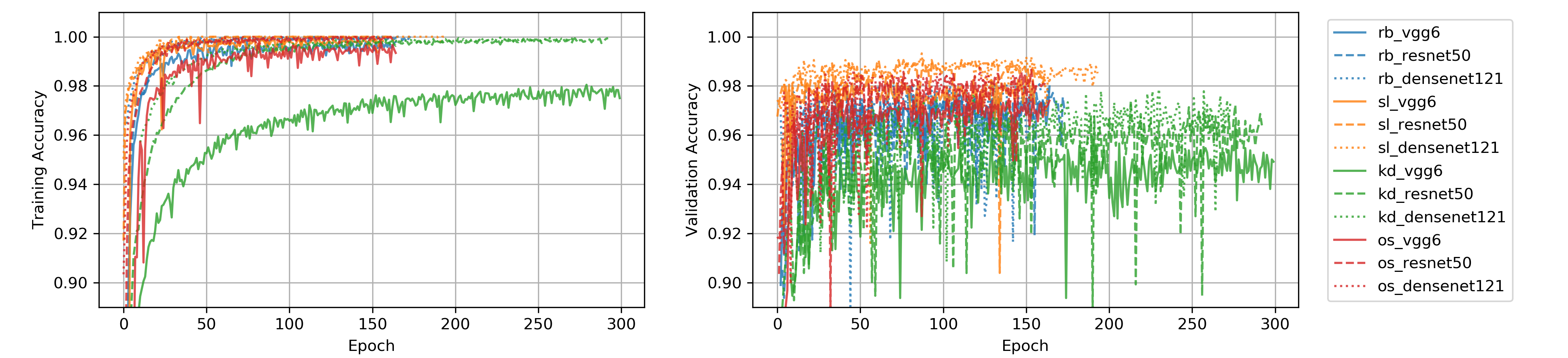}
    \caption{Training (left panel) and validation (right panel) accuracy for all the models that are deployed in production as of January 2019.}
    \label{fig:train_val_acc}
\end{figure*}

\begin{figure*}
  \centering
  \includegraphics[width=0.99\textwidth]{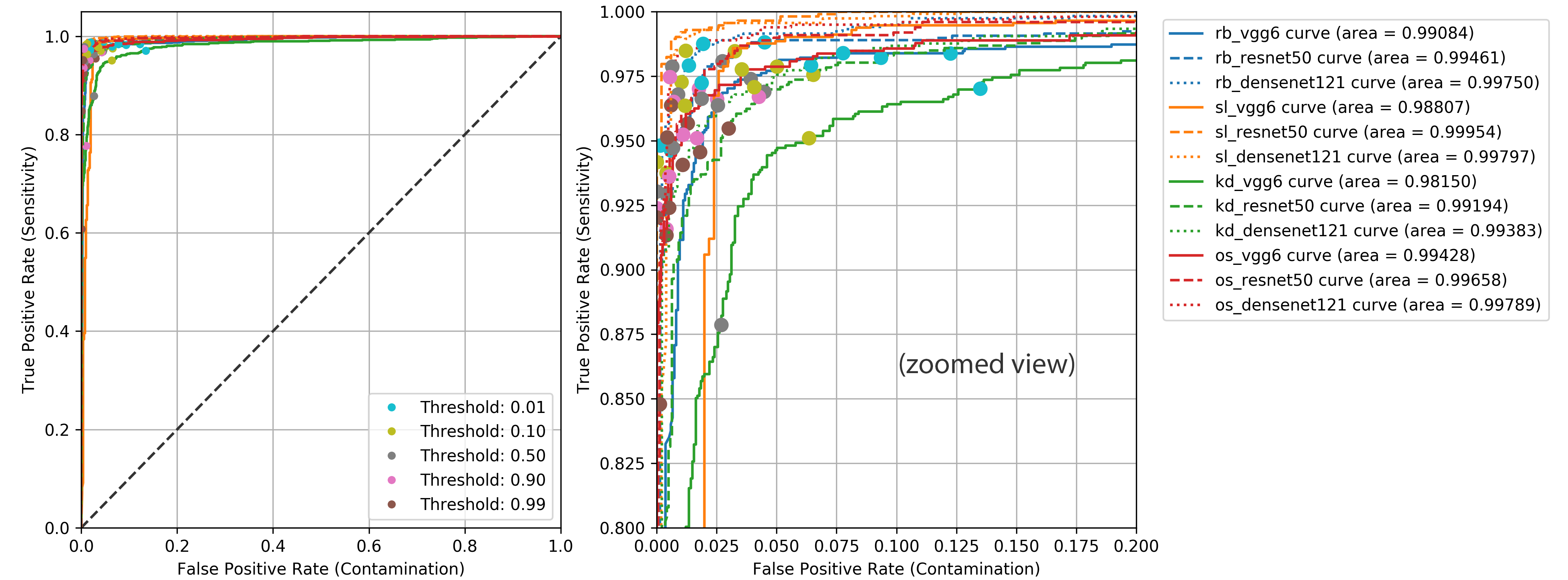}
    \caption{ROC curves for all the models that are deployed in production as of January 2019. The right panel displays a zoomed-in view of the top left corner of the full plot shown on the left panel.}
    \label{fig:roc_rb_sl_kd}
\end{figure*}

\texttt{DeepStreaks} is implemented using \texttt{TensorFlow} software and its high-level \texttt{Keras} API \citep{tensorflow2015-whitepaper, chollet2015keras}. For all models that we trained, we used the binary cross-entropy loss function, the Adam optimizer \citep{2014arXiv1412.6980K}, a batch size of 32, and a 81\%/9\%/10\% training/validation/test data split. The training image data were weighted per class to mitigate the imbalances in the data sets. The images may be flipped horizontally and/or vertically at random. No random rotations and translations were added since those may change the class of the streak for the ``sl'' and ``kd'' classifiers. As it is, the position angles of the streaks adequately sample the range from $0$ to $360$ degrees.

We used the early stopping technique to finish training if no improvement in validation accuracy was observed over many epochs. As a result, the models were trained for 150-300 epochs. For training, we used an on-premise Nvidia Tesla P100 12G GPU. Training a single neural network for 300 epochs on $\sim25k$ images takes about 1.5 hours for the VGG6 architecture, 8 hours for ResNet50, and 10 hours for DenseNet121.

Figure \ref{fig:train_val_acc} shows training (left panel) and validation (right panel) accuracy for all the models that are deployed in production as of January 2019. While the training accuracy for most classifiers reaches over 99\% level after several dozen epochs, the validation accuracy generally stays in the 96-98\% range for the ``rb'' and ``sl'' classifiers, while the ``kd'' classifiers reach 94-97\% validation accuracy. We believe the latter is due to the intrinsic difficulty of the problem to distinguish real short streaks from FMOs observed in excellent (sub-arcsecond) seeing from certain cosmic rays. In our experience, this task is similarly arduous for human scanners.

The test performance of the resulting classifiers as a function of the score threshold is shown on the receiver operating characteristic (ROC) curve, see Fig. \ref{fig:roc_rb_sl_kd}. Evidently, a score threshold of 0.5 that is adopted for all classifiers in \texttt{DeepStreaks} yields 96-98\% true positive rate (TPR) on the test sets while keeping the false positive rate (FPR) low.

\begin{figure}
  \centering
  \includegraphics[width=0.4\textwidth]{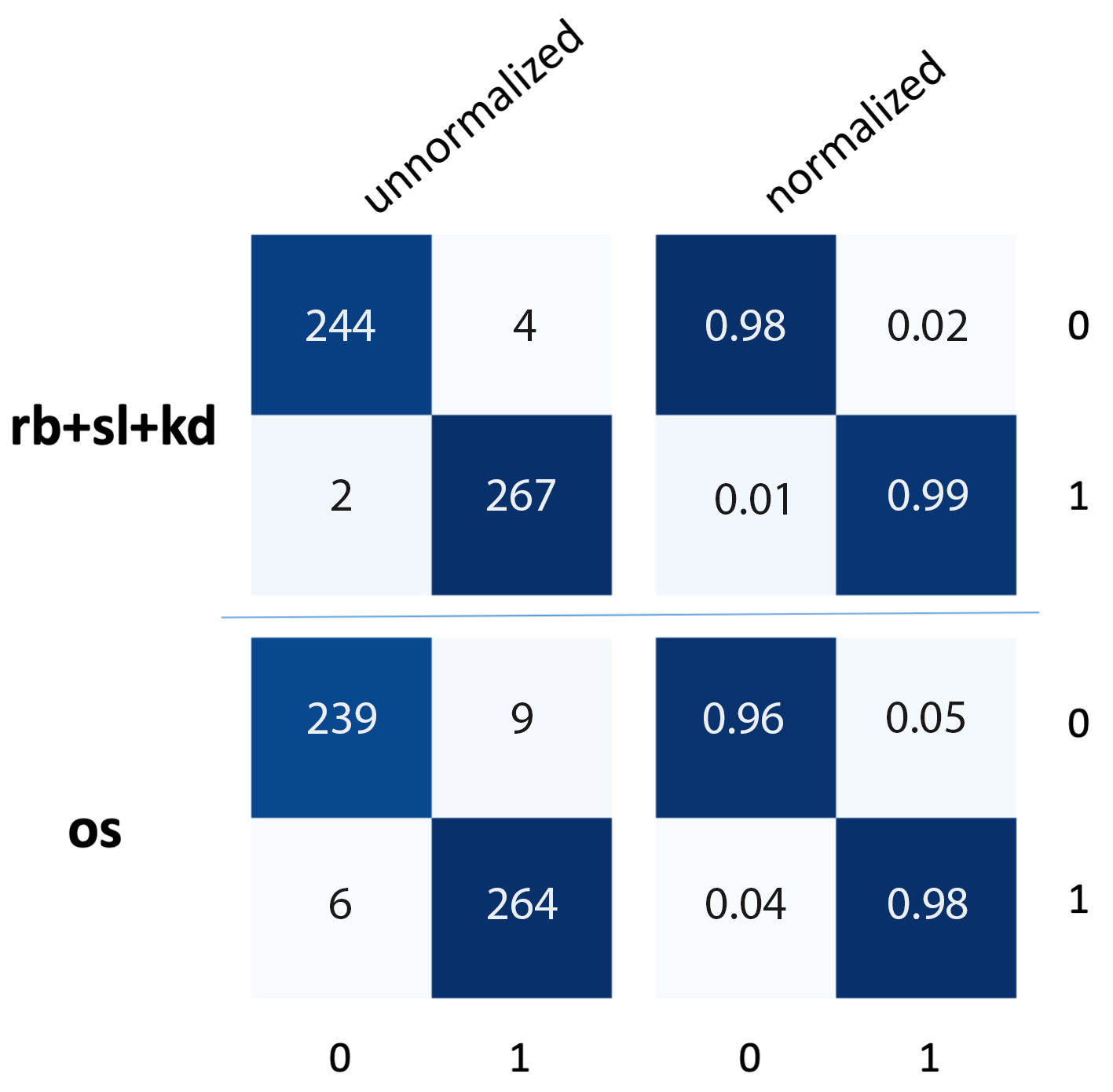}
    \caption{Un-normalized (left column) and normalized (right column) confusion matrices for the ensemble (top row) versus the one-shot (bottom row) classification approach evaluated of a test set of 248 bogus and 270 real streak images from both natural and human-made FMOs. The top-left corner of the matrices shows the number/rate of true negatives, the top-right -- the number/rate of false positives, the bottom-left -- the number/rate of false negatives, and the bottom-right -- the number/rate of true positives.}
    \label{fig:cm_rb_sl_kd__vs__os}
\end{figure}

To assess the performance of the ensemble versus the one-shot classification approach, we constructed a separate test set consisting of 248 bogus and 270 real streak images and evaluated the decision logic depicted in Figures \ref{fig:ds.architecture} and \ref{fig:os.architecture}. As can be seen from the resulting confusion matrices (see Fig. \ref{fig:cm_rb_sl_kd__vs__os}), the approaches show similar performance in terms of precision versus recall on this test set. However, the ensemble (``rb''+``sl''+``kd'') system demonstrated a much better performance when evaluated on all the raw streak cutouts produced by ZTF from December 15, 2018 -- January 15, 2019, which covered a wide range of weather/seeing conditions. Concretely, out of the total of 7 million raw streaks, about 33 thousand (0.5\% of the total) were declared plausible candidates by the ensemble system, whereas the one-shot system output about 8 times more (250 thousand or 3.5\% of the total). Additionally, we ran a sanity check by evaluating the classifiers on a random sample of eight thousand images from the public ImageNet data set. The resulting false positive rate for the ensemble system turned out to be exactly zero, however for the one-shot system, the FPR was around 1\%. For these reasons, \texttt{DeepStreaks} employs the ensemble approach in production.

We chose not to use transfer learning to initialize or freeze layer weights for the deep models and trained all our models from scratch. The reason is that the available pre-trained networks are trained on drastically different image data sets and thus do not necessarily capture the features relevant to this work.

\begin{figure}
  \centering
  \includegraphics[width=0.4\textwidth]{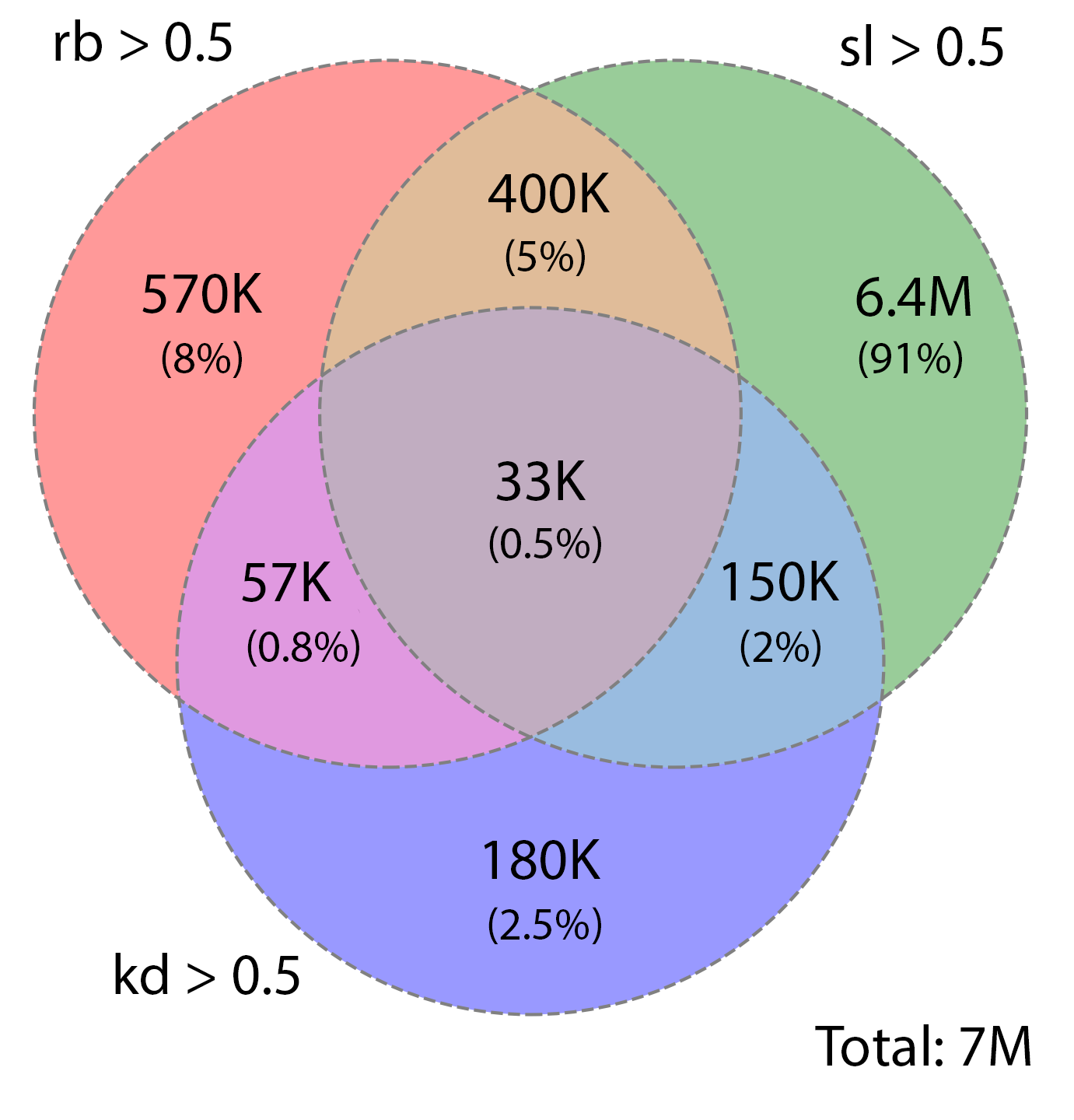}
    \caption{Venn diagram showing the number of streaks that pass \texttt{DeepStreaks}' sub-thresholds and the final number of plausible candidates. ZTF data from December 15, 2018 -- January 15, 2019. ZTF did not observe for 11 nights during that period due to bad weather. The final number of ``good'' candidates output by \texttt{DeepStreaks} (33 thousand) amounts to 0.5\% of the total number of streaks produced by ZTF (compare to 1.7 million, or 25\% of the total output by the RF classifier). 270 streaks out of the 33 thousand plausible candidates were marked as real FMO streaks by human scanners.}
    \label{fig:venn3_rb_sl_kd}
\end{figure}

Figure \ref{fig:venn3_rb_sl_kd} shows the Venn diagram of the number of streaks that pass different \texttt{DeepStreaks}' sub-thresholds and the final number of plausible candidates (see Fig. \ref{fig:ds.architecture}) in ZTF data from December 15, 2018 - January 15, 2019. We note that ZTF did not observe for 11 nights during that period due to bad weather. Human scanners marked 270 out of 33 thousand plausible candidates as real FMO streaks.

While providing a similar sensitivity, \texttt{DeepStreaks} demonstrates a $50\times$ better performance than the original random forest-based classifier used in the ZStreak pipeline in terms of the false negative rate: 1.7 million raw streaks (25\% of the total) were designated plausible candidates by the RF classifier in the same time period. This reduces drastically the time humans have to spend scanning for streaks -- from hours to typically under 10 minutes per day.

\section{Discussion}

The real-time production service that runs \texttt{DeepStreaks} in inference mode is containerized using \texttt{Docker} software. The classifiers are evaluated on batches of raw streak images to utilize vectorization. All individual scores together with the meta-data associated with each streak are saved to a \texttt{MongoDB} NoSQL database\footnote{\url{https://www.mongodb.com/}}. We also built a simple \texttt{flask}\footnote{\url{https://github.com/pallets/flask/}}-based web GUI that provides easy access to the database.

The ZTF FMO Marshal (the web interface of ZStreak) queries the \texttt{DeepStreaks} database every minute and posts \texttt{DeepStreaks}-identified objects in real-time. One or more human scanners then review the stamps on the Marshal and save objects that are potentially interesting for further examination. Compared to the procedure described in \citet{ye2019zstreak}, the introduction of \texttt{DeepStreaks} has reduced the number of stamps posted on the Marshal by $50-100\times$. This vastly reduces the burden on human scanners, and facilitates near-real-time identification of potentially interesting objects.

\begin{figure}
  \centering
  \includegraphics[width=0.4\textwidth]{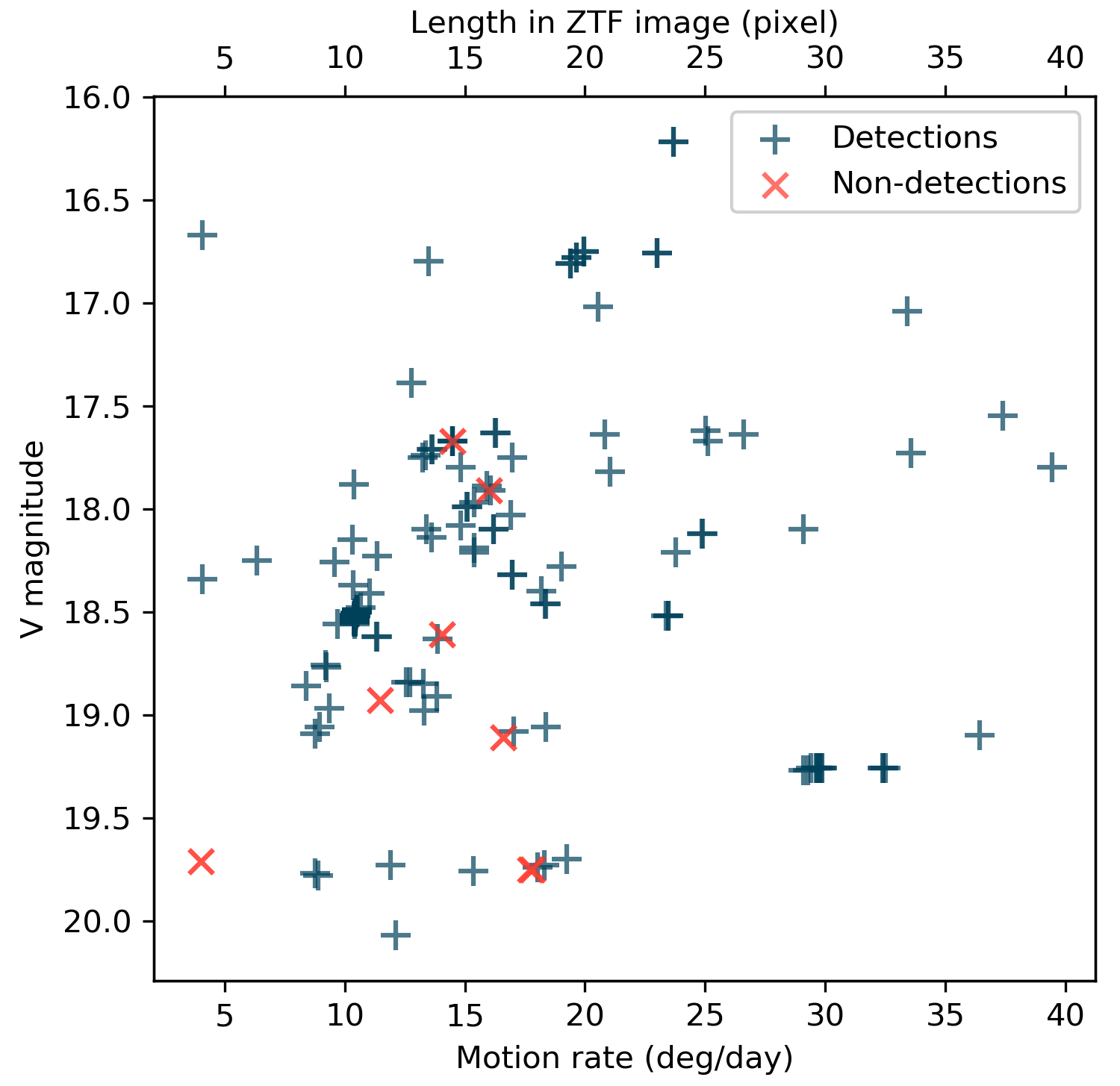}
    \caption{Completeness of \texttt{DeepStreaks} identifications using known NEOs observed by ZTF in October 2018 -- January 2019. Out of 210 streaks from real NEOs detected by the ZTF Streak pipeline at IPAC, 202 (96\%) are correctly classified. ZTF plate scale is $1\arcsec$ per pixel.}
    \label{fig:completeness}
\end{figure}

To quantify the completeness of \texttt{DeepStreaks} identifications, we evaluated it on streak images of known real NEOs detected by the ZTF Streak pipeline from October 2018 -- January 2019 (see Fig. \ref{fig:completeness}). Out of 210 such streaks, 202 (96\%) were correctly classified. We note that the RF classifier initially used in \texttt{ZStreak} with a threshold of 0.05 demonstrates similar TPR on this data set.

We observe a 10-20\% reduction in the number of false positives after each ``dataset enrichment'' campaign that we carried out (see Section \ref{sec:training_performance}). We will continue conducting such campaigns in the future to further reduce the FPR.
The nightly variation in the TPR (96-98\%) appears to be random, however such factors as airmass, seeing or Moon phase etc. may play a role here. It is hard to quantify the effect of these factors at this point due to small number statistics, but we plan to perform a detailed investigation in the future.

\begin{figure*}
  \centering
  \includegraphics[width=0.95\textwidth]{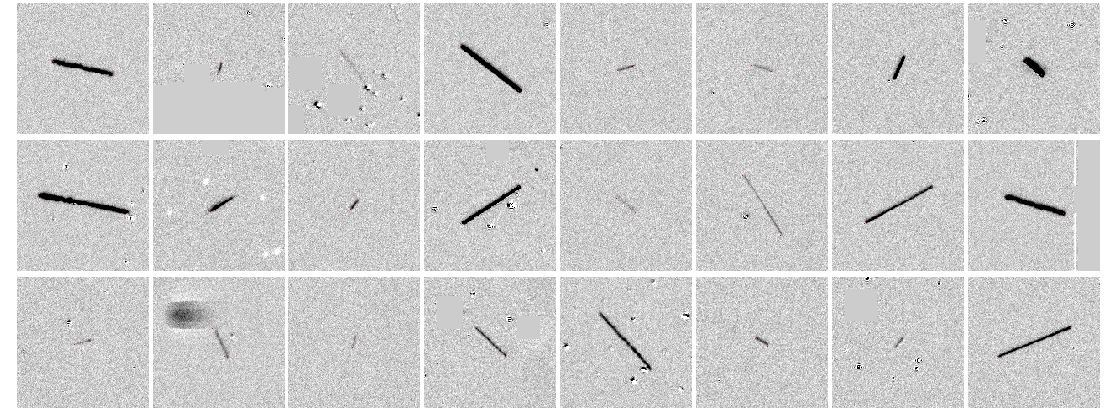}
    \caption{Examples of streaks from real objects, both natural and human-made, identified by \texttt{DeepStreaks} in December 2018 - January 2019. The data were taken under a wide range of weather/seeing conditions (FWHM from $1.5"$ to $4"$).}
    \label{fig:reals_zoo}
\end{figure*}

As of February 1, 2019, 15 NEOs have been discovered with \texttt{DeepStreaks}, including 2019 BE$_5$, the fastest-spinning asteroid discovered to date that has a rotational period of 12 seconds (W. Ryan, private comm.), and 2019 BF$_5$, a PHA. Table \ref{tbl:disc} summarizes the confirmed NEO discoveries. Listed are asteroid designation assigned by the Minor Planet Center, discovery date, $V$-magnitude, apparent motion rate, flyby distance, orbital type and absolute magnitude. Figure \ref{fig:reals_zoo} shows examples of streaks from real fast-moving objects, both natural and human-made, identified by \texttt{DeepStreaks}. The data were taken under a wide range of seeing conditions (FWHM from $1.5"$ to $4"$) and spanned across December 2018 -- January 2019.

\begin{table*}
\begin{center}
\caption{15 confirmed NEOs discovered by \texttt{DeepStreaks} as of February 1, 2019. Listed are designation assigned by the Minor Planet Center, discovery date, $V$ magnitude, apparent motion rate, flyby distance, orbital type and absolute magnitude ($H$). \label{tbl:disc}}
\begin{tabular}{|c|ccc|c|c|c|}
\hline
Prov. des. & \multicolumn{3}{c|}{Disc. circumstance} & Closest dist. & Orbit & $H$ \\
 & Disc. date & $V$ mag & Rate & & & \\
 & & & ($^\circ$/day) & (Lunar Distances) & & \\
\hline
2018 VJ$_{10}$ & 2018 Nov. 15 & 17.5 & 40 & 0.5 & Apollo & 28.6 \\
2018 YM & 2018 Dec. 17 & 19.0 & 35 & 4.0 & Apollo & 27.1 \\
2018 YG$_2$ & 2018 Dec. 16 & 19.2 & 20 & 4.5 & Apollo & 26.0 \\
2018 YO$_2$ & 2018 Dec. 29 & 18.2 & 20 & 0.5 & Apollo & 29.6 \\
2018 YY$_2$ & 2018 Dec. 31 & 18.2 & 20 & 4.5 & Apollo & 25.9 \\
2019 AC$_9$ & 2019 Jan. 10 & 18.3 & 10 & 4.0 & Apollo & 25.7 \\
2019 BZ & 2019 Jan. 24 & 18.6 & 40 & 2.4 & Apollo & 27.5 \\
2019 BU$_2$ & 2019 Jan. 25 & 18.7 & 10 & 9.2 & Apollo & 25.3 \\
2019 BK$_2$ & 2019 Jan. 26 & 18.4 & 30 & 2.8 & Apollo & 26.8 \\
2019 BY$_3$ & 2019 Jan. 28 & 18.9 & 15 & 3.2 & Apollo & 27.4 \\
2019 BL$_4$ & 2019 Jan. 28 & 19.7 & 10 & 2.6 & Apollo & 27.7 \\
2019 BC$_5$ & 2019 Jan. 31 & 19.0 & 15 & 7.0 & Apollo & 25.4 \\
2019 BD$_5$ & 2019 Jan. 31 & 17.7 & 25 & 2.8 & Apollo & 26.7 \\
2019 BE$_5$ & 2019 Jan. 31 & 15.1 & 50 & 3.0 & Aten & 25.0 \\
2019 BF$_5$ & 2019 Jan. 28 & 18.0 & 20 & 9.5 & Apollo & 21.5 \\
\hline
\end{tabular}
\end{center}
\end{table*}

We have demonstrated that by putting together a few simulations, large amounts of data from ZTF, fast computing, and a few deep learning models we can improve the efficiency of detecting streaking asteroids by a couple orders of magnitude, saving tens of human-hours per week at the same time. Our method can be equally easily applied to other data sets, many of which are publicly available. We will continue striving to find fainter objects in ZTF data trying to push for completeness. The additional epochs we gather for known objects will also help build better orbits and hopefully provide early warning should any err in the direction of Earth.

\vspace{\baselineskip}

\texttt{DeepStreaks} code and pre-trained models are available at \url{https://github.com/dmitryduev/DeepStreaks}

\section*{Acknowledgements}
D.A. Duev acknowledges support from the Heising-Simons Foundation under Grant No. 12540303. Q.-Z. Ye is supported by the GROWTH project funded by the National Science Foundation under Grant No. 1545949. Based on observations obtained with the Samuel Oschin Telescope 48-inch Telescope at the Palomar Observatory as part of the Zwicky Transient Facility project. Major funding has been provided by the U.S. National Science Foundation under Grant No. AST-1440341 and by the ZTF partner institutions: the California Institute of Technology, the Oskar Klein Centre, the Weizmann Institute of Science, the University of Maryland, the University of Washington, Deutsches Elektronen-Synchrotron, the University of Wisconsin-Milwaukee, and the TANGO Program of the University System of Taiwan. AM acknowledges support from NSF (1640818 and AST-1815034).

The authors are grateful to Eran Ofek for useful discussions.




\bibliographystyle{mnras}
\bibliography{deepstreaks} 







\bsp	
\label{lastpage}
\end{document}